\documentclass[final]{ws-procs975x65_1}

\IfFileExists{srcltx.sty}{\usepackage[active]{srcltx}}

\usepackage{wrapfig}

\begin{document}

\title{Cosmic connections: from cosmic rays to gamma rays, to cosmic backgrounds and magnetic fields}

\author{Alexander Kusenko\address{Department of Physics and Astronomy, University of California, Los
Angeles, CA 90095-1547, USA 
}%
\address{Kavli IPMU, University of Tokyo, Kashiwa, Chiba 277-8568, Japan 
}
        } 

\begin{abstract}
Combined data from gamma-ray telescopes and cosmic-ray detectors have produced some new surprising insights regarding intergalactic and galactic magnetic fields, as well as extragalactic background light.  We review some recent advances, including a theory explaining the hard spectra of distant blazars and the measurements of intergalactic magnetic fields based on the spectra of distant sources.  Furthermore, we discuss the possible contribution of transient galactic sources, such as past gamma-ray bursts and hypernova explosions in the Milky Way, to the observed flux of ultrahigh-energy cosmic-rays nuclei.  The need for a holistic treatment of gamma rays, cosmic rays, and magnetic fields serves as a unifying theme for these seemingly unrelated phenomena.

\end{abstract}


\bodymatter

\section{Gamma ray astronomy of cosmic rays}

Gamma rays from Active Galactic Nuclei (AGN) are studied extensively using ground-based atmospheric Cherenkov telescopes (ACT), as well as Fermi Space Telescope and other instruments.  Their signals reveal important information about the sources, as well as about extragalactic background light (EBL) and intergalactic magnetic fields (IGMF) along the line of sight.  The same sources are expected to accelerate cosmic rays, 
although it is more difficult to associate cosmic rays with their sources because the local, galactic magnetic fields alter the arrival directions of cosmic rays. 

\subsection{Secondary gamma rays from  the line-of-sight interactions of cosmic rays}
It was recently proposed that the hardness (and uniform redshift-dependent shape) of gamma-ray spectra of distant blazars can be naturally explained by the line-of-sight interactions of cosmic rays accelerated in the blazar jets~\cite{Essey:2009zg,Essey:2009ju,Essey:2010er,Essey:2010nd,Essey:2011wv,Murase:2011cy,Razzaque:2011jc,Prosekin:2012ne}. The cosmic rays with energies below $10^{17}-10^{18}$~eV can cross large distances with little loss of energy and can generate high-energy gamma rays in their interactions with cosmic background photons relatively close to the observer.  Such {\em secondary} gamma rays can reach the observer even if their energies are well above TeV.  In the absence of cosmic-ray contribution, some unusually hard intrinsic spectra\cite{Lefa:2011xh} or hypothetical new particles~\cite{De_Angelis:2007dy,Hooper:2007bq} have been invoked to explain the data.

As long as the IGMFs are smaller than $\sim$10 femtogauss, secondary gamma rays come to dominate the signal from a sufficiently distant source.  One can see this from the way the flux scales with distance for primary and secondary gamma rays~\cite{Essey:2010er}: 

\begin{eqnarray}
F_{\rm primary,\gamma}(d)& \propto &  \frac{1}{d^2} e^{-d/\lambda_\gamma} \label{exponential} 
\end{eqnarray}

\begin{eqnarray}
F_{\rm secondary,\gamma}(d)& \propto &  \frac{\lambda_\gamma}{d^2}\Big(1-e^{-d/\lambda_\gamma}\Big)  \\
& \sim &  \left \{ 
\begin{array}{ll}
1/d, & {\rm for} \ d \ll \lambda_\gamma, \\ 
1/d^2, & {\rm for} \ d\gg \lambda_\gamma .
\end{array} \right.
\end{eqnarray}

Obviously, for a sufficiently distant source, secondary gamma rays must dominate because they don't suffer from the exponential suppression as in Eq.~(\ref{exponential}). The predicted spectrum turns out to be similar for all the distant AGN, depending only on their redshift.  These  predictions are in excellent agreement with the data~\cite{Essey:2009ju,Essey:2009zg,Essey:2010er}.  

\begin{wrapfigure}{r}{0.6\textwidth}
  \begin{center}
      \includegraphics[width=0.55 \textwidth]{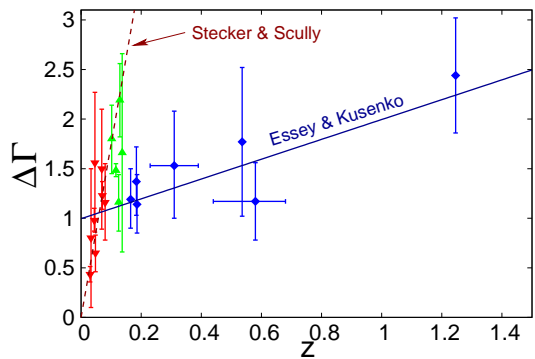}
\caption{
Spectral index change $\delta \Gamma = \Gamma_{\rm GeV} - \Gamma_{\rm TeV}$  as a function of redshift.  While the low-redshift blazars agree with the Stecker--Scully relation\cite{Stecker:2006wq}, the data indicate the existence of an additional, distinct population with a weak redshift dependence at redshifts 0.15 and beyond\cite{Essey:2011wv}.  In particular, the recently measured redshift\cite{Landt:2012it} of PKS 0447-439 is in agreement with the trend.   \label{fig:delta}}
    \end{center}
\end{wrapfigure}

One can see the transition from primary to secondary gamma rays in Fig.~\ref{fig:delta}, which shows the spectral index difference for blazar spectra as a function of their redshifts. 
At small redshifts, the data confirm the Stecker -- Scully relation\cite{Stecker:2006wq}, but, at redshifts 0.15 and beyond, there is clearly a new population of blazars, whose observed spectral index 
shows only a weak dependence on the redshift.  The nearby population is obviously the blazars from which primary gamma rays are observed.  The distant blazars are observed in secondary gamma rays, which are produced in line-of-sight cosmic ray interactions.  These secondary gamma rays are produced relatively close to the observer, regardless of the distance to the source.  Hence, their redshift dependence is much weaker\cite{Essey:2011wv}.  Finally, there is an intermediate population around redshift 1.2 which is composed of some blazars seen in primary gamma rays and some seen in secondary gamma rays. 

A recent redshift measurement of PKS 0447-439 redshift\cite{Landt:2012it} further strengthens our interpretation.  Gamma rays with energies above 1~TeV have been observed from this blazar by HESS\cite{Zech:2011ym}.  The spectral properties agree with the trend (Fig.~\ref{fig:delta}).  Furthermore, there is no way for primary gamma rays to reach Earth from such a distant source, while secondary gamma rays provide a consistent explanation of the PKS 0447-439 spectrum~\cite{Aharonian:2012fu}.

This motivates future observations by ACT of blazars with known large redshifts.  Secondary gamma rays with TeV and higher energies can be observed even from some sources 
located at cosmological ($z\sim 1$) distances~\cite{Aharonian:2012fu}.

\begin{wrapfigure}{r}{0.6\textwidth}
  \begin{center}
      \includegraphics[height=0.45 \textwidth,angle=270]{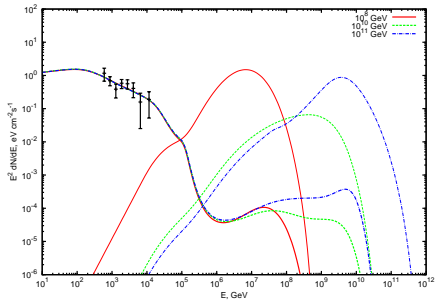}
\caption{Photon (low energy) and neutrino (high energy) spectra~\cite{Essey:2009ju} expected from an AGN at  $z=0.14$ (such as 1ES0229+200), normalized to HESS data points (shown)~\cite{Aharonian:2005gh}, for  $E_{\rm max}=10^8$GeV, $10^{10}$GeV, and $10^{11}$GeV shown by the solid, dashed, and dash-dotted lines, respectively. \label{fig:universality}}
    \end{center}
\end{wrapfigure}

 The spectral slope of protons and the level of EBL do not have a strong effect on the spectrum of secondary photons, as one can see from Fig.~\ref{fig:universality}.  However, for the same photon flux, the neutrino flux varies depending on the maximal energy $E_{\rm max}$ to which the protons are accelerated.  Indeed, there are two competing processes that generate secondary photons: $p\gamma_{EBL}\rightarrow p\pi^0 \rightarrow p\gamma\gamma $ and $p\gamma_{CMB}\rightarrow pe^+e^-$.  For smaller $E_{\rm max}$, a larger fraction of photons come from the hadronic channel, which is accompanied by production of neutrinos via $p\gamma_{EBL}\rightarrow n\pi^{\pm}$ followed by the decays of charged pions and the neutron.  
Neutrino observations can help determine this parameter~\cite{Essey:2009ju}.

\begin{figure}[t!]
\begin{center}
\includegraphics[width=0.48 \textwidth]{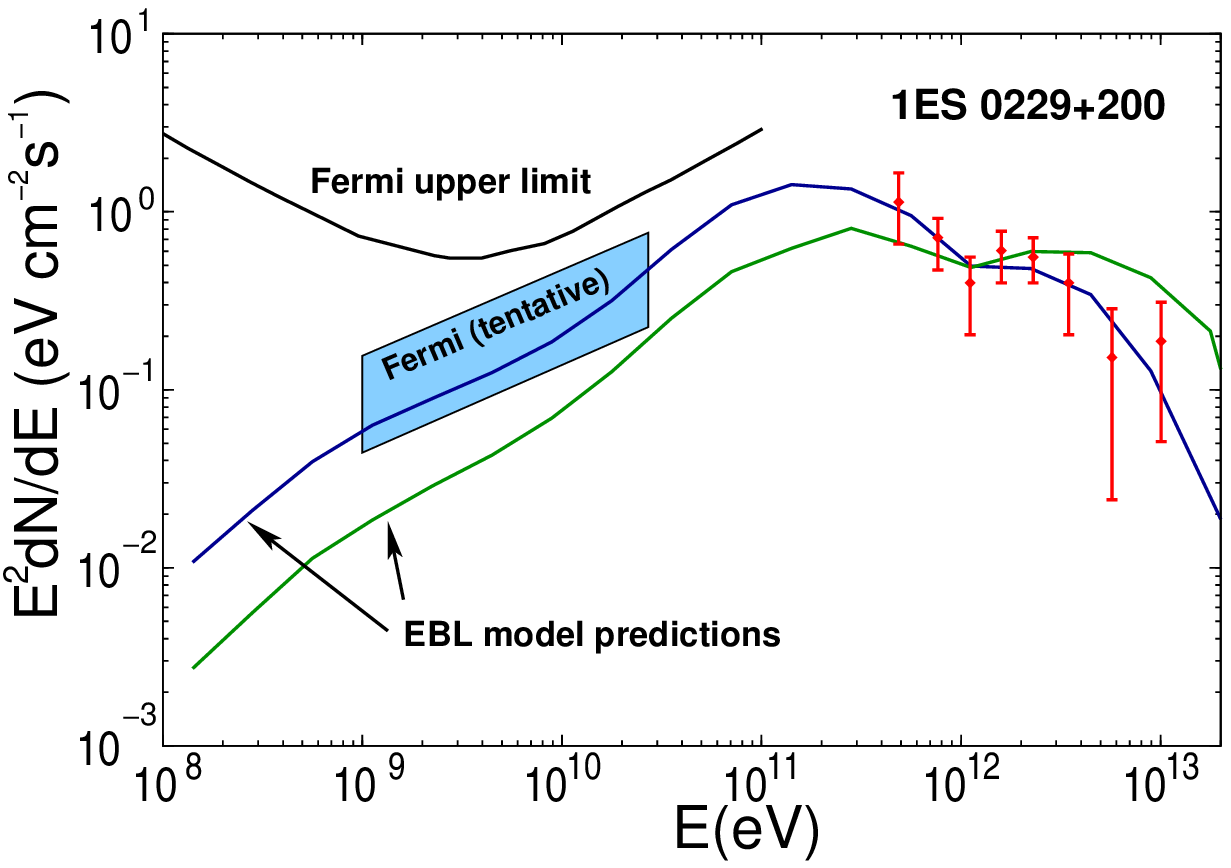} 
\hfill 
\includegraphics[width=0.45 \textwidth]{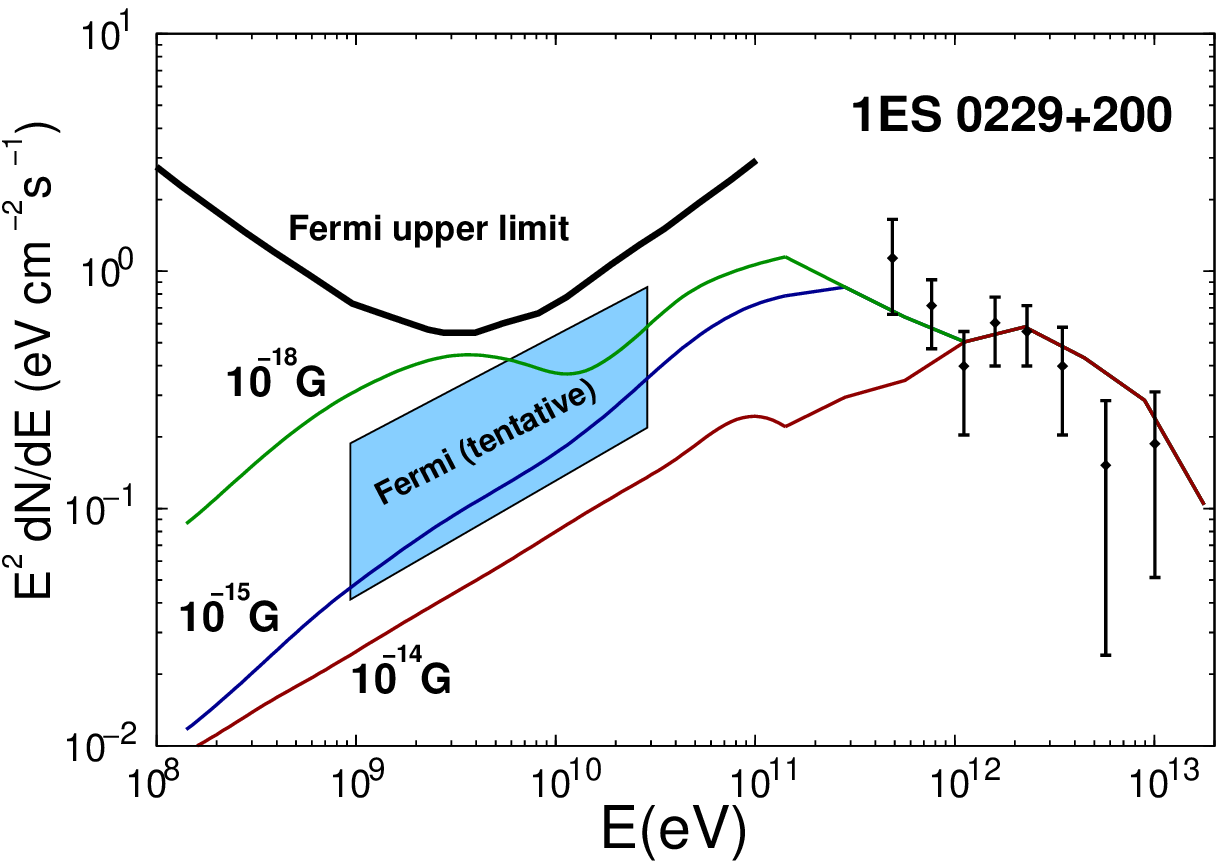}
\caption{Sensitivity of secondary gamma-ray spectra to the model of EBL and to the average value of IGMFs.  {\em Fermi} upper limit and tentative detection~\cite{Orr:2011ha} for blazar 1ES 0229+200 are shown at lower energy, and HESS data points at high energy.  The predictions of two models of EBL are shown in the left panel according to Essey et al.~\cite{Essey:2010er}, and the effects of intergalactic magnetic fields are shown in the right panel~\cite{Essey:2010nd}.  
\label{fig:EBL}}
\end{center}
\end{figure}

\subsection{IGMFs and EBL}
The success of this picture lends support to the hypothesis of cosmic ray acceleration in AGN.  Furthermore, one can use the spectral gamma-ray data to 
study EBL and IGMFs.  The predicted spectra depend to some extent on the EBL model, as shown in Fig.~\ref{fig:EBL}, although this dependence is too weak to distinguish between different models\cite{Essey:2010er}.  IGMFs, however, can have a strong effect on the goodness of fit.  Based on the spectra of several distant blazars, one can set both upper and lower limits on IGMF\cite{Essey:2010nd}: 
$$
10^{-17} {\rm G} < B < 3\times 10^{-14} {\rm G}.
$$

\subsection{Time variability}
\begin{wrapfigure}{l}{0.6\textwidth}
\includegraphics[width=0.6\textwidth,angle=0]{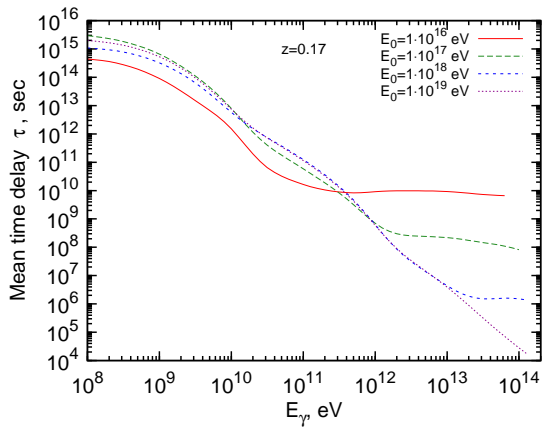}
\caption{\label{fig:delays}  Time delays of gamma rays emitted at redshift $z=0.17$ for different proton energies $E_{0}$ in a femtogauss random IGMF with a correlation length of 1~Mpc.  
}
\end{wrapfigure}

An important property of secondary gamma rays is the lack on short-scale time variability\cite{Prosekin:2012ne}.  For $E>1$~TeV and $z>0.15$, one expects the signal to be dominated by 
secondary photons, and any time variability on short scales should be erased by delays in the propagation of protons and electromagnetic cascades.  Fig.~\ref{fig:delays} shows the time delays
as a function of the proton energy.

The present data are not yet sufficient to probe time variability at the relevant energies and redshifts because the data points are too few.  While time variability has been observed for {\em nearby} TeV blazars at TeV energies, as well as for distant TeV blazars at energies above a few hundred GeV, no variability has been  reported  so far for {\em distant} TeV blazars at {\em TeV  energies}.  
One can infer from Fig.~\ref{fig:delta} how distant the source has to be for the secondary signals to dominate.  It is evident that the secondary component takes over for redshifts beyond 0.15.  

\section{Composition of UHECR and past transient phenomena in the Milky Way}

Let us now turn to another phenomenon related to cosmic rays and magnetic fields, only this time we will concentrate on the magnetic fields inside the galaxy and their effect on 
the observed fluxes of ultrahigh-energy nuclei\cite{Calvez:2010uh,Kusenko:2010dj}.
There is a growing evidence that long GRBs are caused by a relatively rare type(s) of supernovae, while the short GRBs probably result from the coalescence of neutron stars with neutron stars or black holes. Compact star mergers undoubtedly take place in the Milky Way, and therefore short GRBs should occur in our Galaxy. 

Although there is some correlation of long GRBs with star-forming metal-poor galaxies~\cite{Fruchter:2006py}, many long GRBs are observed in high-metallicity galaxies as well~\cite{Savaglio:2006xe,CastroTirado:2007tn,Levesque:2010rn}, and therefore one expects that long GRBs should occur in the Milky Way.  Less powerful hypernovae, too weak to produce a GRB, but can still accelerate UHECR~\cite{Wang:2007ya}, with a substantial fraction of nuclei~\cite{Wang:2007xj,Murase:2008mr}.  

If the observed cosmic rays originate from past explosions in our own Galaxy, PAO results have a straightforward explanation~\cite{Calvez:2010uh}.  

GRBs have been proposed as the sources of extragalactic UHECR~\cite{Waxman:1995vg,Vietri:1995hs,Murase:2008mr}, and they have also been considered as possible Galactic sources~\cite{Dermer:2005uk,Biermann:2003bt,Biermann:2004hi}. It is believed that GRBs happen in the Milky Way at the rate of one per  $t_{\rm GRB}\sim 10^{4}-10^{5}$ years \cite{Schmidt:1999iw,Frail:2001qp,Furlanetto:2002sb,Perna:2003bi,Cui:2010pv}. Such events have been linked to the observations of positrons~\cite{Bertone:2004ek,Parizot:2004ph,Ioka:2008cv,Calvez:2010fd}. 

If local sources, such as past GRBs, hypernovae, and other stellar explosions in the Milky Way, produce a small fraction of heavy nuclei\cite{Horiuchi:2012by}, the observed fraction of UHE nuclei is greatly amplified by diffusion. This  is because the galactic magnetic fields are strong enough to trap and contain nuclei but not protons with energies above EeV.  This observation leads to a simple explanation of the composition trend observed by PAO.

\begin{figure}
\centerline{   \includegraphics[width=.8\textwidth]{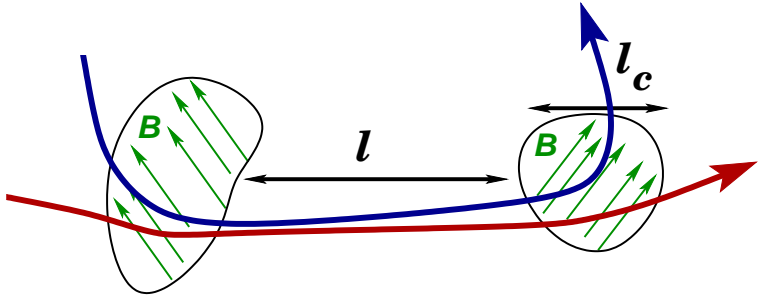}}
  \caption{For each species, there is a critical energy $E_{0,i}$ for which the Larmor radius $R_i$ is equal to the magnetic coherence length $l_c$. For $E\ll E_{0,i}$, the mean free path of the diffusing particle is $l \sim l_0$, and $D_i(E)= l_c/3$.  For $E\gg E_{0,i}$, the particle is deflected only by a small angle $\theta\sim l_0/R_i$, and, after $k$ deflections, the mean deflection angle squared is $\bar{\theta^2}\sim k (l_0/R_i)^2$.  The corresponding diffusion coefficient is $D_i(E) \propto ( \frac{E}{E_{0,i}})^2 , \ {\rm for} \ E \gg  E_{0,i}$. \label{fig:rigidity}}
\end{figure}

As illustrated in Fig.~\ref{fig:rigidity}, diffusion depends on rigidity, and, therefore, the observed composition can be altered by diffusion~\cite{Calvez:2010uh,Wick:2003ex}. Changes in composition due to a magnetic fields have  been discussed in connection with the spectral ``knee''~\cite{Wick:2003ex}, and also for a transient source of UHECR~\cite{Kotera:2009ms}.  The ``knee'' in the spectrum occurs at lower energies than those relevant PAO, and at higher energies the cosmic rays effectively probe the spectrum of magnetic fields on greater spatial scales, of the order of 0.1~kpc~\cite{Han:2004aa}.  

One can use a simple model~\cite{Calvez:2010uh} to show how diffusion affects the observed spectrum of the species ``$i$'' with different rigidities.  Let us suppose that all species are produced with the same spectrum $n_i^{(\rm src)}=n_0^{(\rm src)}\propto E^{-\gamma} $ at the source located in the center of the Milky Way and examine the observed spectra altered by the energy dependent diffusion and by the  trapping in the Galactic fields.

 In diffusive approximation, the transport inside the Galaxy can be described by the equation:

\begin{eqnarray}
\frac{\partial n_i}{\partial t} - \vec{\nabla}(D_i \vec{\nabla}n_i)  +\frac{\partial}{\partial E}(b_i n_i )= \nonumber \\
 Q_i(E,\vec{r},t) +\sum_k\int P_{ik}(E,E') n_k(E') dE'. \nonumber
\end{eqnarray}

Here $D_i(E,\vec{r},t)=D_i(E)$ is the diffusion coefficient, which we will assume to be constant in space and time. The energy losses and all the interactions that change the particle energies
are given by $b_i(E)$ and the kernel in the collision integral $P_{ik}(E,E')$. For energies below GZK cutoff, one can neglect the energy losses on the diffusion time scales.

The diffusion coefficient $D(E)$ depends primarily on the structure of the magnetic fields in the Galaxy. Let us assume that the magnetic field structure is comprised of uniform randomly oriented domains of radius $l_0$ with a constant field $B$ in each domain. The density of such domains is $N\sim l_0^{-3}$. The Larmor radius depends on the particle energy $E$ and its electric charge $q_i=eZ_i$:
\begin{eqnarray}
R_i & =&   l_0 \left( \frac{E}{E_{0,i}} \right),\ {\rm where} \
E_{0,i}= E_0 \, Z_i, \\
E_0 & = &  10^{18} {\rm eV} \left(\frac{B}{3\times 10^{-6}\, {\rm G}}\right) \left(\frac{l_0}{0.3\, {\rm kpc}} \right)   . \label{E0i}
\end{eqnarray}

The spatial energy spectrum of random magnetic fields inferred from observations suggests that $B\sim 3\mu {\rm G} $ on the 0.3~kpc spatial scales, and that there is a significant change at $l=1/k\sim 0.1-0.5$~kpc~\cite{Han:2004aa}. This can be understood theoretically because the turbulent energy is injected into the interstellar medium by supernova explosions on the scales of order 0.1~kpc.  This energy is transferred to smaller scales by direct cascade, and to larger scales by inverse cascade of magnetic helicity.  Single-cell-size models favor $\sim 0.1$~kpc scales as well~\cite{Han:2004aa}.

As explained in the caption of Fig.~1, diffusion occurs in two different regimes depending on whether the Larmor radius is small or large in comparison with the correlation length.  As a result, the diffusion coefficient changes its behavior dramatically at $E=E_{0,i} $:

 \begin{equation}
 D_i(E) = \left \{
\begin{array}{ll}
D_0 \left ( \frac{E}{E_{0,i}} \right)^{\delta_1}, & E \le  E_{0,i}, \\
D_0 \left ( \frac{E}{E_{0,i}} \right)^{(2-\delta_2)},  & E > E_{0,i} .
\end{array} \right.
\end{equation}

Here the two parameters $0 \le \delta_{1,2} \le 0.5$ are different from zero if the magnetic domains are not of the same size. The exact values of 
these parameters depend on the power spectrum of turbulent magnetic fields. 

The approximate solution of the transport equation in our simple model yields 
\begin{equation}
n_i(E,r) = \frac{Q_0}{4\pi r\, D_i(E)} \left ( \frac{E_0}{E} \right )^\gamma.
\label{solution_GC}
\end{equation}
\begin{wrapfigure}{r}{0.6\textwidth}
  \includegraphics[width=0.6\textwidth]{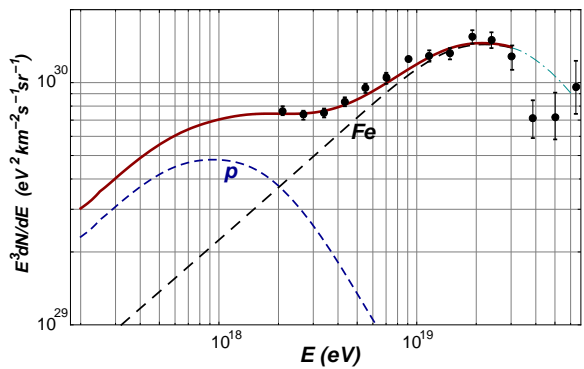}
  \caption{UHECR spectra~\cite{Calvez:2010uh}, for the magnetic field $\sim 10\mu$G, coherent over $l_0=100$~pc domains. The power and the iron fraction were adjusted to fit the PAO data~\cite{Abraham:2009wk}. \label{fig:UHECR_spectra}
}
\end{wrapfigure}
Since diffusion depends on rigidity, the composition becomes energy dependent.  Indeed, at critical energy $E_{0,i}$, which is different for each nucleus, the solution (\ref{solution_GC}) changes from $\propto E^{-\gamma} $ to $\propto E^{-\gamma-2} $ because of the change in $D_i(E)$, as discussed in the caption of Fig.~1.  Since the change occurs at a rigidity-dependent critical energy $E_{0,i}=e E_0 Z_i$, the larger nuclei lag behind the lighter nuclei in terms of the critical energy and the change in slope.  If protons dominate for $E<E_0$, their flux drops dramatically for $E>E_0$, and the heavier nuclei dominate the flux.  The higher $Z_i$, the higher is the energy at which the species experiences a drop in flux.

The model~\cite{Calvez:2010uh} provides a qualitative description of the data (see Fig.~\ref{fig:UHECR_spectra}). 
To reproduce the data more accurately, it must be improved.  First, one should use a more realistic source population model.  Second, one should include the coherent component of the Galactic magnetic field.  Third, one should not assume that UHECR comprise only two types of particles, and one should include a realistic distribution of nuclei. Finally, one should include the extragalactic component of UHECR produced by distant sources, such as active galactic nuclei (AGN) and GRBs (outside the Milky Way).  A recent realization that very high energy gamma rays observed by Cherenkov telescopes from distant blazars are likely to be secondary photons produced in cosmic ray interactions along the line of sight lends further support to the assumption that cosmic rays are copiously produced in AGN jets~\cite{Essey:2009zg,Essey:2009ju}. For energies $E>3\times 10^{19}$~eV, the energy losses due to photodisintegration, pion production, pair production and interactions with interstellar medium become important and must be included. The propagation distance in the Galaxy exceeds 10~Mpc, so that the Galactic component should exhibit an analog of GZK suppression in the spectrum.  Extragalactic propagation can also affect the composition around $10^{18}$~eV~\cite{Hill:1983mk}.

\begin{figure}
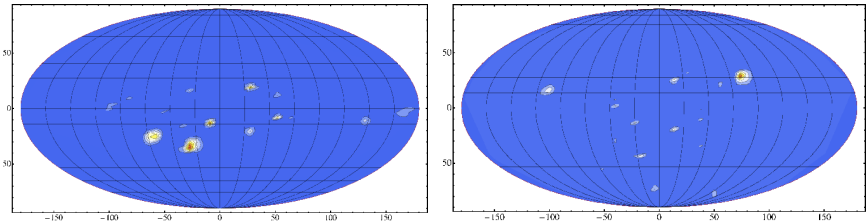

\centerline{
\resizebox{0.45\columnwidth}{!}{%
\includegraphics{uhecr_skymap_contour.eps2}} 
\resizebox{0.45\columnwidth}{!}{\includegraphics{uhecr_shGRB_skymap_contour.eps2}}}
\caption{Monte Carlo simulation of a typical set of ''hot spots'' in the directions of the closest, recent GRBs or hypernovae, in the model of Calvez {\em et al.}~\cite{Calvez:2010uh}.  The distribution of transient sources was modeled for long GRBs (left) and short GRBs (right). \label{fig:hot_spots} }
\end{figure}

Galactocentric anisotropy for a source distribution that traces the stellar counts in the Milky Way is small~\cite{Calvez:2010uh}. Although the anisotropy in protons is large at high energies, their contribution to the total flux is small, so the total anisotropy was found to be $<10\%$, consistent with the observations.  The latest GRBs do not introduce a large degree of anisotropy, as it would be in the case of UHE protons, but they can create ``hot spots'' and clusters of events (Fig.~\ref{fig:hot_spots}).

The model~\cite{Calvez:2010uh} leads to the following prediction for the highest-energy cosmic rays.  Just as the protons of the highest energies escape from our Galaxy, they should escape from the host galaxies of remote sources, such as AGN. Therefore, UHECR with $E>3\times 10^{19}$~eV should correlate with the extragalactic sources.  Moreover, these UHECR should be protons, not heavy nuclei, since the nuclei are trapped in the host galaxies.  If and when the data will allow one to determine composition on a case-by-case basis, one can separate $E>3\times 10^{19}$~eV events into protons and nuclei and observe that the protons correlate with the nearby AGN.  This prediction is one of the non-trivial tests of our model: at the highest energies the proton fraction should exist and should correlate with known astrophysical sources, such as AGN.  The microgauss magnetic fields in the Milky Way cause relatively small deflections for the highest-energy protons.  As for the intergalactic magnetic fields, there are reasons to believe that they are relatively weak, of the order of a femtogauss~\cite{Essey:2010nd}, and, therefore, they should not affect the protons significantly on their trajectories outside the clusters of galaxies.   

If local, Galactic GRBs are the sources of UHECRs, the energy output in cosmic rays should be of the order of $10^{46}$~erg per GRB.  This is a much lower value than what would be required of extragalactic GRBs to produce the same observable flux.  Indeed, in our model the local halo has a much higher density of UHECR than intergalactic space, and so the overall power per volume is much smaller.  The much higher energy output required from extragalactic GRBs~\cite{Waxman:1995vg,Vietri:1995hs,Murase:2008mr} in UHECR has been a long-standing problem.  The same issue does not arise in our case because it seems quite reasonable that a hypernova or some other unusual supernova explosion would generate $10^{46}$~erg of UHECR with energies above 10~EeV.

\section{A gamma-ray signature of cosmic-ray nuclei}

A spectral feature, namely an ``iron shoulder`` at 5-10~GeV can help identify cosmic nuclear accelerators\cite{Kusenko:2011tb}.  Nuclei are likely to come out of acceleration regions unstable because 
they can lose a nucleon or a few nucleons to photodisintegration in the high-density photon environments accompanying some accelerators\cite{Horiuchi:2012by}.  An unstable nucleus decays, and most of such decays are $\beta$-decays.  With a probability of order one, the $\beta$-decay electron is captured by the Coulomb potential of the fully ionized atom\cite{Bahcall:1961zz}.  Hence, a non-negligible fraction of nuclei come out of astrophysical accelerators in the form of one-electron ions.   
\begin{wrapfigure}{r}{0.7\textwidth}
  \includegraphics[width=0.7 \textwidth]{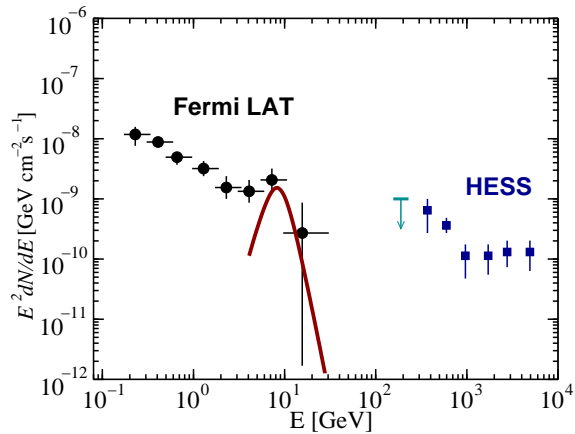}
  \caption{ Expected signature of nuclear emission for Cen A (solid line), normalized to total observed flux of iron, 
and the data from Fermi~\cite{Falcone:2010fk} and HESS~\cite{Aharonian:2009xn,Aharonian:2005ar}.\label{signature}
}
\end{wrapfigure}

In a narrow energy range, CMB photons have energies $\approx 7 $~keV in the rest frame of the ion.  Such photons can excite the ion, which later emits a 7~keV photon (in the ion's rest frame).  Multiple excitations and de-excitations can take place resulting in emission of gamma rays, which have energies of $5-10$~GeV in the laboratory frame. 
The spectral feature around 8~GeV (Fig.~\ref{signature}) can be used for identifying astrophysical sources of nuclei, or (in the case of non-detection) for setting the upper limits on nuclear acceleration\cite{Kusenko:2011tb}.

\section{Conclusions}

Based on the recent data, one can make several remarkable inferences about the ultrahigh-energy cosmic rays and magnetic fields inside the Milky Way and in the intergalactic space. 
Gamma-rays detected from most distant blazars are most likely dominated by the secondary photons produced in line-of-sight interactions of cosmic rays.  This interpretation allows one to 
set both upper and lower bounds on intergalactic magnetic fields,  $10^{-17} {\rm G} < B < 3\times 10^{-14} {\rm G}$\cite{Essey:2010nd}. 

 Furthermore, the energy dependent composition of UHECR, with heavier nuclei at high energy, points to a non-negligible contribution from Galactic sources~\cite{Calvez:2010uh}.  Diffusion in turbulent Galactic magnetic field traps the nuclei more efficiently than protons, leading to an increase in the nuclear fraction up to the energy at which iron escapes ($\sim 30$~EeV).   At higher energies, the extragalactic protons should dominate the flux of UHECR, and their arrival directions should correlate with locations of the known sources.  

If and when the neutrino telescopes, such as IceCube~\cite{Halzen:2010yj}, detect point sources, one can learn about the cosmic-ray sources and photon backgrounds by comparing the neutrino flux to the photon flux. Neutrino and gamma-ray observations can help distinguish the local Galactic sources from extragalactic sources of UHE nuclei~\cite{Murase:2010gj,Murase:2010va,Hooper:2010ze}. These inferences open exciting new opportunities for multi-messenger photon, charged-particle, and neutrino astronomy.  

This work was supported by DOE Grant DE-FG03-91ER40662.

\bibliographystyle{ws-procs975x65}
\bibliography{cosmic}

\end{document}